\documentclass[usenatbib,usegraphicx]{mn2e}
\usepackage{color}
\usepackage{aas_macros}
\newcommand\ha{H$\alpha$~}
\newcommand\hb{H$\beta$~}
\newcommand\oii{[O~{\sc ii}]~}
\newcommand\oiii{[O~{\sc iii}]~}

\title[]{A large scale structure traced by [O~{\sc ii}] emitters hosting a distant cluster at $z$=1.62}
\author[K. Tadaki et al.]
{Ken-ichi Tadaki$^{1}$\thanks{E-mail:tadaki.ken@nao.ac.jp}, 
Tadayuki Kodama$^{2}$, Kazuaki Ota$^{3}$, Masao Hayashi$^{4}$, 
\newauthor 
Yusei Koyama$^{4}$,
Casey Papovich$^{5,6}$,
Mark Brodwin$^{7}$,
Masayuki Tanaka$^{8}$,
\newauthor
Masanori Iye$^{1,4}$
\newauthor \\
$^{1}$Department of Astronomy, Graduate School of Science, University of Tokyo, Tokyo 113-0033, Japan\\
$^{2}$Subaru Telescope, National Astronomical Observatory of Japan, 650 North Afohoku Place, Hilo, HI 96720, USA\\
$^{3}$Department of Astronomy, Kyoto University, Kitashirakawa-Oiwake-cho, Sakyo-ku, Kyoto 606-8502, Japan\\
$^{4}$Optical and Infrared Astronomy Division, National Astronomical Observatory of Japan, Mitaka, Tokyo 181-8588, Japan\\
$^{5}$Department of Physics and Astronomy, Texas A\&M University, College Station, TX 77845-4242, USA\\
$^{6}$George P. and Cynthia Woods Mitchell Institute for Fundamental Physics and Astronomy\\
$^{7}$Department of Physics, University of Missouri, 5110 Rockhill Road, Kansas City, MO 64110\\
$^{8}$Institute for the Physics and Mathematics of the Universe, University of Tokyo, Kashiwanoha, Kashiwa, Chiba 277-8582, Japan\\
}
\begin{document}

\date{Accepted . Received ; in original form }

\pagerange{\pageref{firstpage}--\pageref{lastpage}} \pubyear{}

\maketitle

\label{firstpage}

\begin{abstract}
We present a panoramic narrow-band imaging survey of \oii emitters in and around the ClG J0218.3--0510 cluster at $z$=1.62 with Suprime-Cam on Subaru telescope. 352 \oii emitters were identified on the basis of narrow-band excesses and photometric redshifts. We discovered a huge filamentary structure with some clumps traced by \oii emitters and found that the ClG J0218.3--0510 cluster is embedded in an even larger super-structure than the one reported previously. 
31 \oii emitters were spectroscopically confirmed with the detection of \ha and/or \oiii emission lines by FMOS observations.
In the high density regions such as cluster core and clumps, star-forming \oii emitters show a high overdensity by a factor of more than 10 compared to the field region. 
Interestingly, the relative fraction of \oii emitters in photo-$z$ selected sample does not depend significantly on the local density.
Although the star formation activity is very high even in the cluster core, some massive quiescent galaxies also exits at the same time.
Furthermore, the properties of the individual \oii emitters, such as star formation rates, stellar masses and specific star formation rates, do not show a significant dependence on the local density, either. 
Such lack of environmental dependence is consistent with our earlier result by \cite{2011MNRAS.415.2670H} on a $z=1.5$ cluster and its surrounding region.
The fact that the star-forming activity of galaxies in the cluster core is as high as that in the field at $z\sim1.6$ may suggest that the star-forming galaxies are probably just in a transition phase from a starburst mode to a quiescent mode, and are thus showing comparable level of star formation rates to those in lower density environments.
We may be witnessing the start of the reversal of the local SFR--density relation due to the ``biased'' galaxy formation and evolution in high density regions at high this redshift, beyond which massive galaxies would be forming vigorously in a more biased way in proto-cluster cores.
\end{abstract}


\section{Introduction}
\label{sec;intro}

It is widely known that the formation and evolution of galaxies strongly depend on their surrounding environments. Galaxy clusters are most massive object and the densest structures in the universe. As such, clusters and their surrounding regions serve as ideal sites for studying the roles of galaxy environment on galaxy formation and evolution. In the local universe, it is known that the star formation has already ceased long time ago in high density regions such as clusters and most of the galaxies therein are only passively evolving since then, while galaxies in the general field are still actively forming stars even at the present day (e.g., \citealt{2002MNRAS.334..673L,2003ApJ...584..210G,2004MNRAS.353..713K}). Such environmental variation has been well recognized as a sharp correlation between galaxy morphology and local number density of galaxies (e.g., \citealt{1980ApJ...236..351D,1984ApJ...281...95P,1993ApJ...407..489W,1997ApJ...490..577D}). 
In clusters at $z\sim1$, \cite{2011arXiv1112.3655M} found that the specific star formation rate (SSFR) of star-forming galaxies is independent of environment at fixed stellar mass but the fraction of star-forming galaxies is decreased in high density environment. This suggests that the environmental-quenching timescale would be rapid. 
Recent observations are discovering very distant clusters at $z\sim 1.5-2.0$ \citep{2011ApJ...732...33B,2011A&A...527L..10F, 2010ApJ...716.1503P,2010ApJ...716L.152T,2010ApJ...725..615H,2011A&A...526A.133G}. 
It is crucial to observe clusters at such frontier redshifts so as to directly witness and understand the early evolution of galaxies and the physical processes which are responsible for the strong environmental dependence of galaxy properties at later times.

To survey star-forming galaxies with a narrow-band filter, that captures nebular emission lines such as \ha and \oii, is very efficient and effective for studying cluster galaxies because we can construct a large sample of star-forming galaxies without any expensive spectroscopic observation. 
The correlation between the star formation activity, traced by a nebular emission, and the environment has already been investigated by many authors rather intensively. 
For field environment and groups at $z=$0.84, \cite{2011MNRAS.411..675S} have constructed a large sample of \ha emitters with a narrow-band filter as a part of the High-z Emission Line Survey (HiZELS), and found that the fraction of star-forming galaxies falls sharply as a function of local number density of photo-$z$ sample from about 40 percent in the field to almost zero in rich groups. For clusters at $z \sim$ 0.5--1.5, our previous studies on \ha emitters in CL0939+4713 (Abell 851) at $z=0.41$ and in RX J1716.4+6708 at $z=0.81$ show that the fraction of emitters decreases towards cluster central region, and the star formation activity peaks in the intermediate density regions such as groups and filaments at the cluster outskirts \citep{2011ApJ...734...66K, 2010MNRAS.403.1611K}. 
On the other hand, the studies in higher redshift clusters ($z\sim$ 1.5) have produced two different results. While there are no \ha emitters within a radius of 200 kpc from the centre of the most massive cluster XMMU J2235.3$-$2557 at $z=1.39$ \citep{2011MNRAS.411.2009B}, the other cluster XMMXCS 2215.9$-$1738 at $z=1.46$ \citep{2010MNRAS.402.1980H} shows much higher star formation activity as traced by \oii emitters even in the cluster core.
One possible interpretation would be that, in a more massive system galaxy evolution is accelerated and star formation is truncated at an earlier epoch.
However, there is an caveat that some of the \oii\ emitters in the cluster core may be
contaminated by AGN.

To obtain more general picture on the star formation history in clusters, we need to study many clusters at each epoch. In fact, we are currently conducting the MAHALO-Subaru project (MApping HAlpha and Lines of Oxygen with Subaru: Kodama et al. in prep), which is a large and systematic narrow-band survey of star forming galaxies in many clusters at $z\sim$ 1.5--3.0 and in the field.
This paper reports one of the initial results of the project.
This survey will provide us with the dependence of star formation history on environment.

We comment here, however, that if we only target clusters that have already formed and matured at each epoch, the results may be biased towards apparently weaker evolution.
This difficult problem should be tackled by targeting lower-mass systems as well as the rich clusters. By tracing the surrounding large scale structures around rich clusters, however, we can automatically include many smaller mass groups and/or filaments embedded in the
structures. This is another good advantage of panoramic studies of distant clusters
like ours.

In this paper, we report a narrow-band survey of \oii emitters in and around the spectroscopically confirmed distant cluster ClG J0218.3-0510 (IRC 0218-A) at $z=1.62$ \citep{2010ApJ...716.1503P,2010ApJ...716L.152T} and is in Subaru/$XMM-Newton$ Deep Survey Field (SXDS; \citealt{2008ApJS..176....1F}). The coordinates of this cluster is $\alpha=2^h18^m21.3^s, \delta=-5^\circ 10\arcmin27\arcsec$(J2000), derived from the centroid of the sources selected by $Spitzer$ IRAC colour in this overdensity. We assume cosmological parameters of H$_0$ = 70 km s$^{-1}$ Mpc$^{-1}$, $\Omega _\mathrm{M}$ = 0.3, and $\Omega _\Lambda$ = 0.7, and adopt AB magnitudes throughout this paper. 
At $z=1.62$, 1\arcsec  corresponds to 8.47 kpc in a physical distance and to 22.20 kpc in a comoving distance.

\section{Observations and data}

\begin{figure}
\centerline{\includegraphics[width=1.0\linewidth]{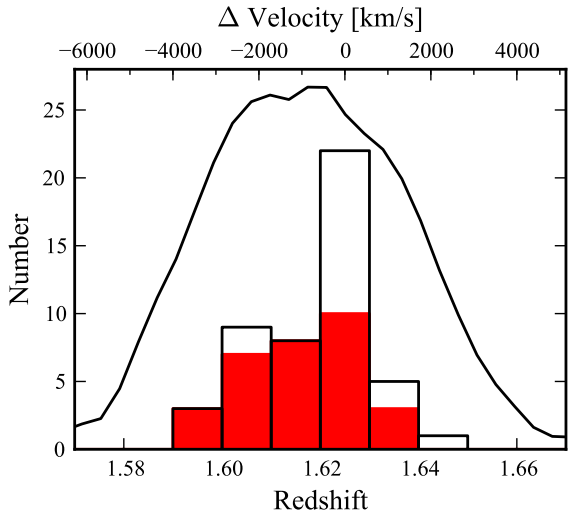}}
\caption{A transmission curve (solid line) of NB973 filter and the redshift distribution (black histogram) of spectroscopically confirmed galaxies, including the cluster members in the literatures \citep{2010ApJ...716.1503P,2010ApJ...716L.152T}. 
The \oii emitters, whose redshifts are derived from our FMOS observation, are shown by red histogram.
}
\label{fig;profile}
\end{figure}

\subsection{Imaging data}

To identify line emitters, we used a narrow-band filter (NB973) and a medium-band filter ($z_R$). 
The NB973 filter ($\lambda _\mathrm{c}=9755$\AA, $\Delta \lambda_\mathrm{FWHM}=202$\AA) can detect an \oii line emission at $1.590\leq z\leq1.644$ and $z_R$ filter ($\lambda _\mathrm{c}=9860~$\AA, $\Delta \lambda_\mathrm{FWHM}=590$\AA) can measure a continuum flux at same wavelength range as the NB973 filter. These filters enable us to sample most of the star-forming galaxies associated to the cluster at $z=1.62$ as shown in Figure \ref{fig;profile}. The imaging data with NB973 filter was obtained in an area between the SXDS-C and SXDS-S by \cite{2010ApJ...722..803O}. The seeing size in the combined image is 1.0\arcsec~and the 5$\sigma$ limiting magnitude is 25.4 in a 2.0\arcsec~diameter aperture. The details of the observations and the data reduction are described in \cite{2010ApJ...722..803O}. For the $z_R$ data, we conducted a new observation with Subaru Prime Focus Camera (Suprime-Cam; \citealt{2002PASJ...54..833M}) on the Subaru telescope in October 2010. The weather was good during our observation, and the sky conditions were photometric with the seeing of 0.5 - 0.7\arcsec.
Data reduction was carried out in a normal manner with SDFRED \citep{2002AJ....123...66Y,2004ApJ...611..660O}. The frames for the total of 327 minutes are integrated in the final image. The field coverage is 830 arcmin$^2$, corresponds to $1.4\times 10^5$ Mpc$^3$ in the survey volume.  The 5$\sigma$ limiting magnitude is 25.3 in a 2.0\arcsec~diameter aperture. Point spread function (PSF) of the $z_R$ image was degraded to 1.0\arcsec~to match the quality of the NB973 image.

\subsection{Photometric catalogue}

In the SXDS field, many multi-wavelength data are already available. In this work, we used a total of 12 multi-band data including the NB973 and $z_R$-band ones. Deep optical broadband images ($B,V,R,i',z'$) were taken by the Suprime-Cam \citep{2008ApJS..176....1F}. For near-infrared wavelength, the UKIDSS Ultra Deep Survey (UKIDSS-UDS;\citealt{2007MNRAS.379.1599L}) has been conducted in this region, and we used the DR8 data ($J,H,K$). The photometric
system and calibration are described by \cite{2006MNRAS.367..454H} and \cite{2009MNRAS.394..675H}, respectively.
From 10-band (NB973,$B,V,R,i',z',z_\mathrm{R},J,H,K$) images, we made a photometric catalogue using SExtractor \citep{1996A&AS..117..393B}. PSFs of all the images were matched to 1.0\arcsec and the positions were matched with respect to the NB973 image. A source detection was performed on the NB973 image. The extraction criteria were at least 9 pixels with fluxes above 2$\sigma $ level, where 1$\sigma $ is the sky noise of the image. The sources fainter than 5$\sigma$ limiting magnitude in NB973 were rejected. Photometries in all the images were carried out using the double image mode.
An aperture magnitude within a diameter of 2\arcsec was used to derive a colour and Kron magnitude was used to calculate a total magnitude \citep{1980ApJS...43..305K}.
For mid-infrared wavelength, we used IRAC catalogue (only 3.6 $\mu$m and 4.5 $\mu$m) for a $Spitzer$ Public legacy survey of the UKIDSS-UDS (SpUDS; PI: J. Dunlop). From our catalogue, we identified sources that have an association in the IRAC catalogue within a 1.0 arcsec radius.
Moreover, not only the NB973-detected catalogue but also the K-detected one was created above 5$\sigma$ in the same manner when we select passively evolving galaxies later in section 3.3.
The NB973-detected and the K-detected catalogues include 39229 objects and 40786 objects, respectively, of which 33242 objects are in common.

\subsection{Spectroscopy}

Near-infrared spectroscopy of our photometrically identified \oii emitter candidates was conducted with Fiber Multi Object Spectrograph (FMOS; \citealt{2010PASJ...62.1135K}) on the Subaru telescope in January 2012. Observations were made with low-resolution (LR) mode in IRS1 and high resolution (HR) mode at J-long and H-long bands in IRS2. 
Low-resolution, J-long and H-long modes can cover the wavelength ranges of 1.0--1.8 $\mu $m, 1.12--1.35 $\mu$m and 1.59--1.80 $\mu $m, respectively. Cross-beam switching mode was adopted and the total integration times were 375 minutes in LR, 60 minutes in J-long and 300 minutes in the H-long mode.
The seeing was typically 1\arcsec\ in R-band during observations.
Reduction was carried out with the FMOS pipeline FIBRE-pac \citep{2011arXiv1111.6746I}. The flux calibration was done using some F,G or K-type stars from the Two Micron All Sky Survey (2MASS; \citealt{2006AJ...131.1163S}), observed simultaneously. 
The flux loss was estimated to roughly 50\% for a point source and this might be caused by various factors (e.g. weather condition, instrument focus, and position error in the catalogue). In this paper, only spectroscopic redshifts, derived from fitting emission lines, are used in order to estimate the accuracy of the photometric redshift, and the level of completeness and contamination. Other analyses such as line fluxes, line widths and the line ratios will be presented in detail in our forthcoming paper.
Also, we used the catalogues of spectroscopic redshifts from \cite{2012MNRAS.tmp.2493S} and \cite{2008MNRAS.389..407S}. 

\section{Target selection}
\label{sec;target}

In order to investigate galaxy evolution, we need to sample both actively star-forming galaxies and passively evolving galaxies at the same epoch. As noted in section \ref{sec;intro}, the former can be selected by the presence of nebular emission lines such as \oii. The latter population can be selected by utilizing the photometric redshift and the colour-colour diagram. Using the multi-band data in section~2, a total of 352 \oii emitters at $z\simeq 1.62$ have been identified over a 830 arcmin$^2$ area. 
Also, we have constructed a photo-$z$ selected sample with M$_\mathrm{star}>10^{10}$ M$_\odot$ at $1.56 \leq z_\mathrm{phot} \leq 1.68$, which includes 132 \oii emitters and 259 quiescent galaxies at $z\simeq 1.6$.

\subsection{\oii emitters}
\label{sec;oii}

\begin{figure}
\centerline{\includegraphics[width=\linewidth]{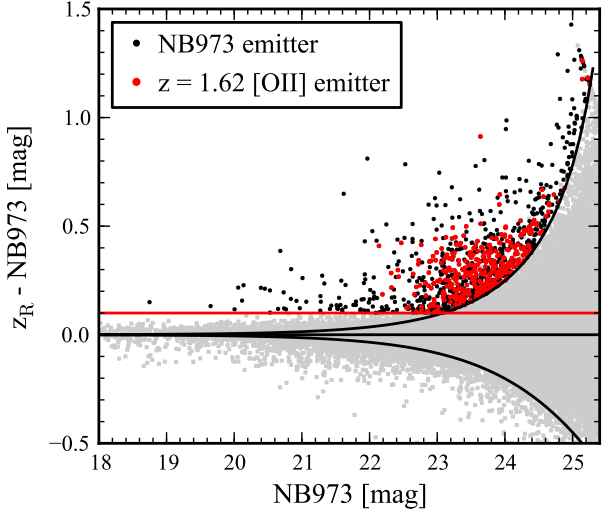}}
\caption{$z_R$ - NB973 colour-magnitude diagram to select \oii emitter candidates at z=1.62. Gray dots show all objects with $\mathrm{NB973} > 5\sigma$. Black and red filled circles are NB973 emitters and \oii emitters which satisfy the criteria in section 3.1.}
\label{fig;nb-bb}
\end{figure}

We can identify emission line galaxies on the basis of the excesses of NB973 fluxes over $z_R$ fluxes. 
NB973-detected catalogue created in section 2.2 is used.
Figure \ref{fig;nb-bb} shows the $z_R$ $-$ NB973 colour-magnitude diagram for all the objects satisfying $\mathrm{NB973} > 5\sigma$.
Note that we made a correction of +0.11 in the $z_R$ $-$ NB973 colour because the effective wavelengths of two filters are not exactly same.
For faint objects ($<2\sigma$) in $z_R$-band, $2\sigma$ value was used to calculate colour.
If an strong emission line comes into the $z_R$ band but not on the NB973 filter (e.g., at $z=1.58$ or $z=1.66$), such object may be detected as false-absorbers.
Also, because the effective wavelength of NB973 filter is slightly shorter than that of the $z_R$ filter, very red objects may have negative colours in $z_R$- NB973.
\cite{1995MNRAS.273..513B} have defined the significance of the excess in the narrow-band by the parameter $\Sigma$ with taking account of the fact that the fainter the narrow-band flux is, the larger the photometric error is. $\Sigma$=2.5 was adopted as our first criteria to select NB emitters \citep{2009MNRAS.398...75S}. Since the standard deviation of $z_R$ $-$ NB973 is 0.03 in the range of 18$<$NB973$<$21, the colour excess of $z_R$ $-$ NB973 $> 0.1$, that is corresponds to  3$\sigma $ for the bright objects, was also adopted. These criteria correspond to the limiting line flux of $1.8\times 10^{-17}$ erg s$^{-1}$ cm$^{-2}$, that is 4-5 M$_\odot$yr$^{-1}$ in dust-uncorrected star formation rate (SFR) using the calibration of \cite{1998ARA&A..36..189K}, and to the equivalent width of about 30 \AA~ in the observed frame. 
Based on first criteria, 969 objects are identified as NB973 emitters. 

\begin{figure}
\centerline{\includegraphics[width=1.0\linewidth]{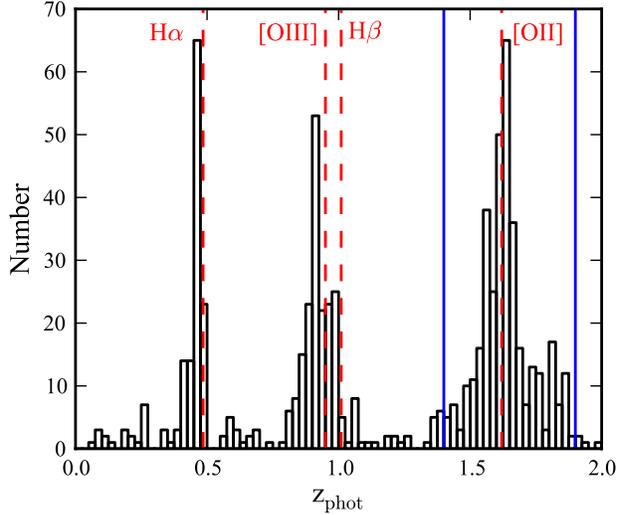}}
\caption{The distribution of photometric redshift for NB973 emitters. Red dash lines indicate the expected redshift of emission lines that can fall onto the NB973 filter (H$\alpha$ at $z\sim0.49$, \oiii at $z\sim0.95$, H$\beta$ at $z\sim1.01$ and \oii at $z\sim1.62$). Blue lines indicate the photometric redshift ranges ($1.4< z_{\mathrm{phot}}< 1.9$) to define \oii emitters.}
\label{fig;photz}
\end{figure}

The NB973 filter can pick out not only \oii lines at $z=1.62$ but some other lines at different redshifts: Ly $\alpha $ at $z=7.02$, H$\beta $ at $z=1.01$, \oiii at $z=0.95$ and H$\alpha $ line at $z=0.49$. To discriminate between \oii emitters and those other line emitters, we utilize photometric redshifts for those objects that have satisfied our criteria. Photometric redshifts were determined with the EAZY code \citep{2008ApJ...686.1503B}. 11 band data ($B, V, R, i', z', z_R, J, H, K, 3.6 \mu \mathrm{m}, 4.5 \mu \mathrm{m}$) are used for SED fitting. 
Capturing Balmer/4000\AA ~break, $z_R$-band should improve the accuracy of photometric redshift for galaxies at $z\sim$1.4--2.0. 
Because $z_R$-band can be substantially affected by the presence of our target emission line itself, the emission line flux and the emission line subtracted continuum flux density are estimated from the $z_R$-band and NB973-band fluxes as follows;

\begin{equation}
F_\mathrm{line}=\Delta _\mathrm{NB973} \frac{f_\mathrm{NB973}-f_{z_R}}{1-\Delta _\mathrm{NB973} /\Delta _{z_R}},
\end{equation}
\begin{equation}
f_\mathrm{continuum}=\frac{f_{z_R}-f_\mathrm{NB973}(\Delta _\mathrm{NB973}/\Delta _{z_R})}{1-\Delta _\mathrm{NB973}/\Delta _{z_R}},
\end{equation}

\noindent
where $F_\mathrm{line}$ is the emission line flux, $f$ denotes a flux density, and $\Delta$ indicates a FWHM of a filter.
The emission line subtracted flux density in $z_R$-band is used for SED fitting.
Figure \ref{fig;photz} is the distribution of the photometric redshift for NB973 emitters. We can clearly recognize three peaks in the photometric redshift distribution (i.e., $z\sim$ 0.5, 1, and 1.6). These redshift peaks neatly correspond to H$\alpha $, \oiii or H$\beta $, and \oii respectively, assuring the detection of line emitters as expected. This also indicates that our first criterion of $\Sigma$=2.5 and $z_R$ $-$ NB973 $> 0.1$ is effective in identifying secure line emitters. 
Out of NB973 emitters, we down selected 352 \oii emitters that fall between $1.4< z_{\mathrm{phot}}<1.9$, as second criterion, to distinguish between our target \oii emitters at $z=1.62$ and other unwanted line emitters at other redshifts.

While most of the \oii emitters are likely to be star-forming galaxies, there is a possibility that some of the \oii emission lines are originated from central active galactic nuclei (AGN) rather than star forming regions. In particular, some authors reported that ``red'' \oii emitters tend to be contaminated by AGN activities \citep{2006ApJ...648..281Y, 2010ApJ...716..970L, 2011MNRAS.415.2670H, 2011arXiv1110.0979T}. Alternatively, red emitters are dusty starburst galaxies which are reddened by a large amount of dust extinction. In any case, these red emitters are interesting objects which may be key populations to understanding galaxy evolution in dense environment. However, we can not disentangle these two possibilities at this stage without spectroscopy. With spectroscopy, we should be able to distinguish between AGNs and dusty starbursts by measuring emission line ratios of [O~{\sc iii}]/\hb and [N~{\sc ii}]/\ha and plot them on the BPT diagram \citep{1981PASP...93....5B,2003MNRAS.346.1055K}. Here, we defined red emitters with the $z - J$ versus $J - K$ colour-colour diagram (hereafter $zJK$ diagram) and treated only the \oii emitters that are outside the quiescent zone defined in section~3.3 as star-forming galaxies. 
Also, our criteria of \oii emitters select three X-ray detected objects \citep{2008ApJS..179..124U}, of which two are located at different redshifts.
Therefore, the remaining one is not used in our analysis due to some uncertainty.
Eventually, we have identified totally 340 star-forming \oii emitters and 12 red emitters associated to the cluster at $z=1.62$.

\subsection{Spectroscopic follow-up for \oii emitters}

\begin{figure*}
\centerline{\includegraphics[width=\linewidth]{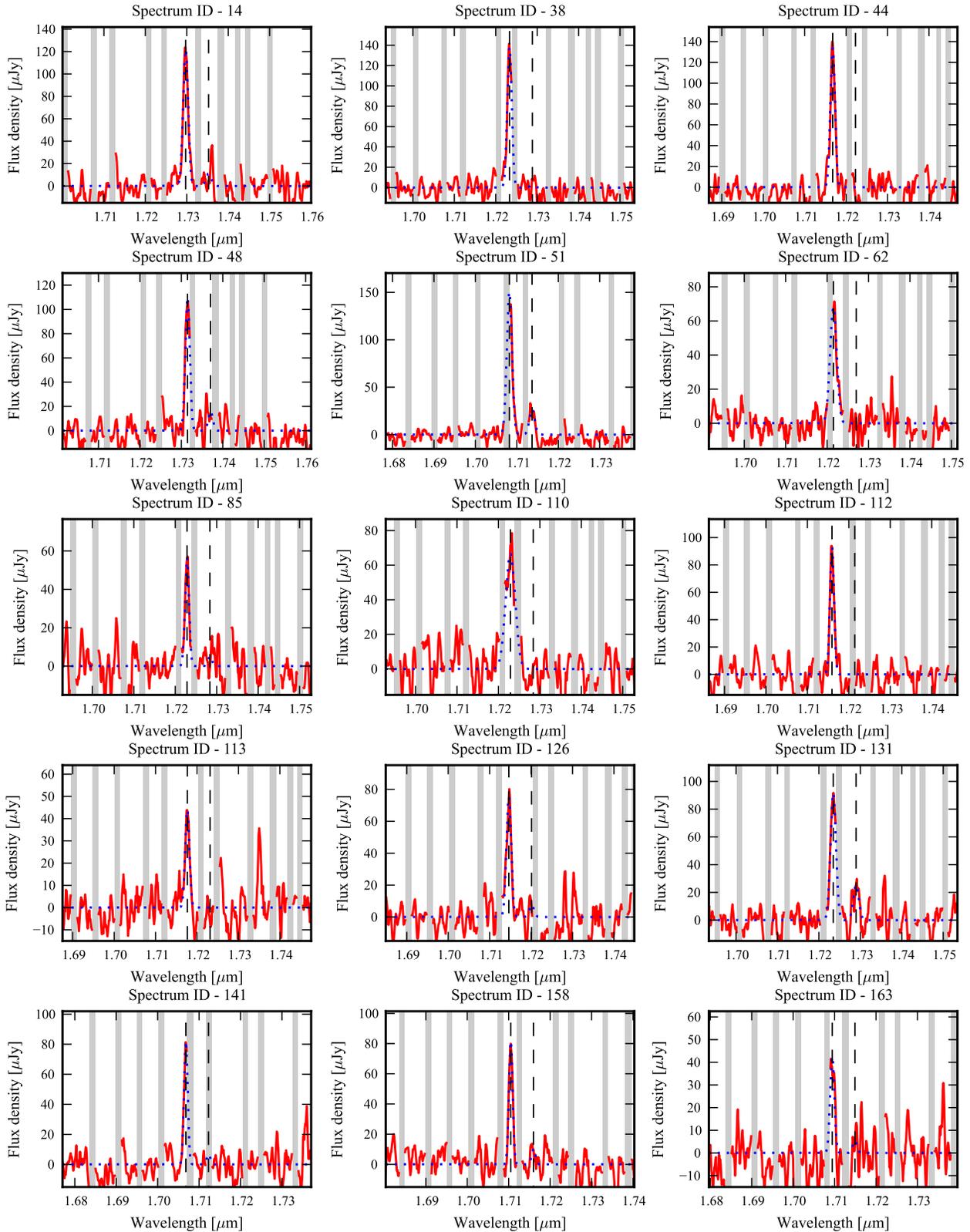}}
\caption{Examples of \ha spectra for \oii emitters. Red line and blue dotted line shows a observed spectra and a gaussian fitting spectra, respectively. Vertical dashed lines indicate the redshifted wavelength of \ha and [N~{\sc ii}]$\lambda6584$. Gray zones indicate the positions of the OH-mask.}
\label{fig;spectra}
\end{figure*}

At $z=1.62$, \oiii and \ha emission lines shift to 1.31 $\mu$m and 1.72 $\mu$m, respectively.
From the NB973 emitter sample with F([O~{\sc ii}])$>3.0\times 10^{-17}$ erg s$^{-1}$ cm $^{-2}$ identified in the previous section, 46 (IRS1, LR mode) and 39 objects (IRS2, HR mode) were observed with FMOS. 
\ha emission lines are detected for 4 objects in LR mode and for 24 in H-long. \oiii emission lines are detected for 7 objects in LR mode and for 4 in J-long. 
The sensitivity of the LR mode is not good enough to detect emission lines due to the large flux loss.
If only the H-long mode is considered, the detection rate of emission lines is about 60 percent in total and 80 percent for bright \oii emitters with F([O~{\sc ii}])$>7.0\times 10^{-17}$ erg s$^{-1}$ cm $^{-2}$.
These value are really good since 20-25 percent of the wavelength coverage is masked in the FMOS spectrograph due to OH-airglow lines.
Figure \ref{fig;spectra} shows some examples of \ha spectra of our \oii emitters.
Spectroscopic redshifts are derived by gaussian fitting with free parameters of redshift, line width, and flux densities of \ha and [N~{\sc ii}]$\lambda6584$ or [O~{\sc iii}]$\lambda4959,\lambda5007$.
As a result, 31 spectroscopic redshifts are obtained in total from the FMOS data, since there are 8 objects whose \ha and \oiii lines are both detected.
There are a total of 42 spectroscopically confirmed galaxies with emission lines, including 11 spectroscopic redshifts with a emission line in the literatures. 40 out of them were selected by the criteria in the previous section. On the other hand, there were no objects with a different redshift in our \oii emitter sample.

\begin{figure}
\centerline{\includegraphics[width=\linewidth]{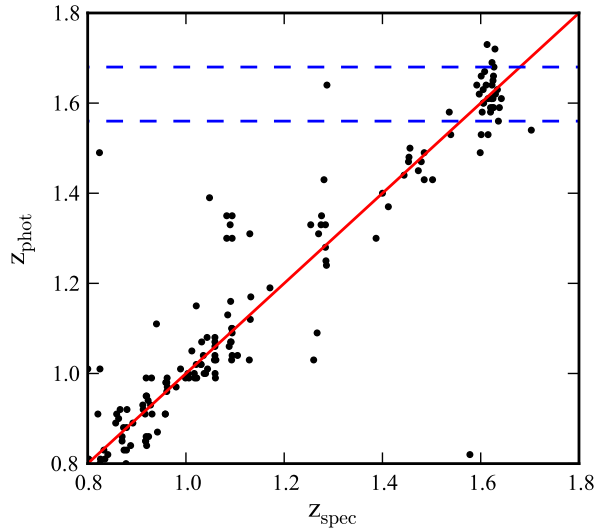}}
\caption{The spectroscopic redshift versus the photometric redshift for galaxies with $0.8<z_{\mathrm{spec}}<1.8$. Red line is $z_\mathrm{phot}=z_\mathrm{spec}$. Blue dashed lines show the criterion of the photo-$z$ selected sample ($1.56\leq z_{\mathrm{phot}}\leq1.68$).}
\label{fig;specz}
\end{figure}

\subsection{Quiescent galaxies}
\label{sec;red}

We also need to select passively evolving galaxies located at $z$=1.62 to quantify the averaged star formation activity.
First, photometric redshifts were measured for all the objects in the K-band detected catalogue created in section~2.
X-ray detected objects were rejected because the accuracy of photometric redshift is so bad that could increase contaminations.
In Figure \ref{fig;specz}, photometric redshifts are plotted against spectroscopic redshifts for confirmed galaxies. The standard deviation of ($z_\mathrm{spec}-z_\mathrm{phot}$) is 0.05 at $1.590\leq z_\mathrm{spec}\leq 1.644$. Our photometric redshifts are fairly good for them. We selected only the objects that fall within a narrow redshift interval of $1.56\leq z_{\mathrm{phot}}\leq 1.68$ since the K-band selected galaxies can include contaminations at contiguous redshift unlike the NB973 emitters.
Such a stringent criterion of photometric redshift range is reasonable
because our photometric redshifts are good enough to recover most of the spectroscopically confirmed cluster members.
In fact the photometric redshift distribution of the \oii emitters shows a narrow concentration at $z\sim 1.6$ as shown in Figure \ref{fig;photz}.

\begin{figure}
\centerline{\includegraphics[width=\linewidth]{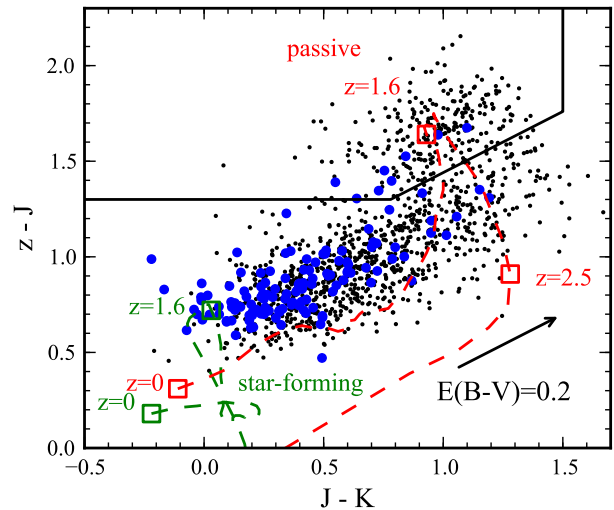}}
\caption{The $z - J$ versus $J - K$ colour-colour ($zJK$) diagram of the photo-$z$ selected galaxies (black dots). Blue circles show the \oii emitter samples. Red dashed and green dashed lines shows a model track of a passive galaxy and a star-forming galaxy, respectively \citep{1998A&A...334...99K}.
The squares indicate the model colour at $z=0, z=1.6$ and $z=2.5$. The arrow shows the dust reddening vector of E(B-V)=0.2 \citep{2000ApJ...533..682C}. The solid lines show the adopted criteria to separate between quiescent galaxies and star-forming galaxies at $z=1.6$.}
\label{fig;quiescent}
\end{figure}

We have constructed a mass-limited sample with the stellar mass M$_\mathrm{star}>10^{10}$ M$_\odot$ on the basis of photometric redshifts.
Stellar masses are estimated from total K-band magnitudes, $K^{\mathrm{total}}$, and $z-K$ colours, using the ratio between mass-to-luminosity ratio in K-band and z-K colour of the population synthesis bulge-disk composite models \citep{1999MNRAS.302..152K} after it is scaled to the Salpeter IMF for consistency.
At $z=1.6$, we used the following equations:

\begin{equation}
\mathrm{log}(M_\mathrm{star}/10^{11})=-0.4(K^{\mathrm{total}}-21.64)
\end{equation}
\begin{equation}
\Delta \mathrm{log}M_\mathrm{star}=0.1-1.12\exp [-0.82(z'-K)].
\end{equation}

\noindent
The amount of contamination and that of completeness of the sample are estimated based on the spectroscopically confirmed galaxies.
There are 32 galaxies with M$_\mathrm{star}>10^{10}$ M$_\odot$ which are located within $1.590\leq z_\mathrm{spec}\leq 1.644$.
Since 6 spectroscopically confirmed cluster member galaxies do not fall within the photometric redshift range of $1.56\leq z_\mathrm{phot}\leq 1.68$, the completeness can be estimated to be 81\%.
We include them in our sample.
Also, because our sample include three objects at different spectroscopic redshift, from the cluster redshift, the contamination can be estimated to be 10\%.
However, in order to accurately estimate them, we need to make more spectroscopic follow-up observations.
132 \oii emitters with M$_\mathrm{star}>10^{10}$M$_\odot$ satisfy our photometric redshift selection ($1.56\leq z_\mathrm{phot}\leq 1.68$).
After excluding these \oii emitters from our photo-$z$ selected sample,
we constructed a non-emitter sample of 1174 galaxies.

\begin{figure*}
\centerline{\includegraphics[width=\linewidth]{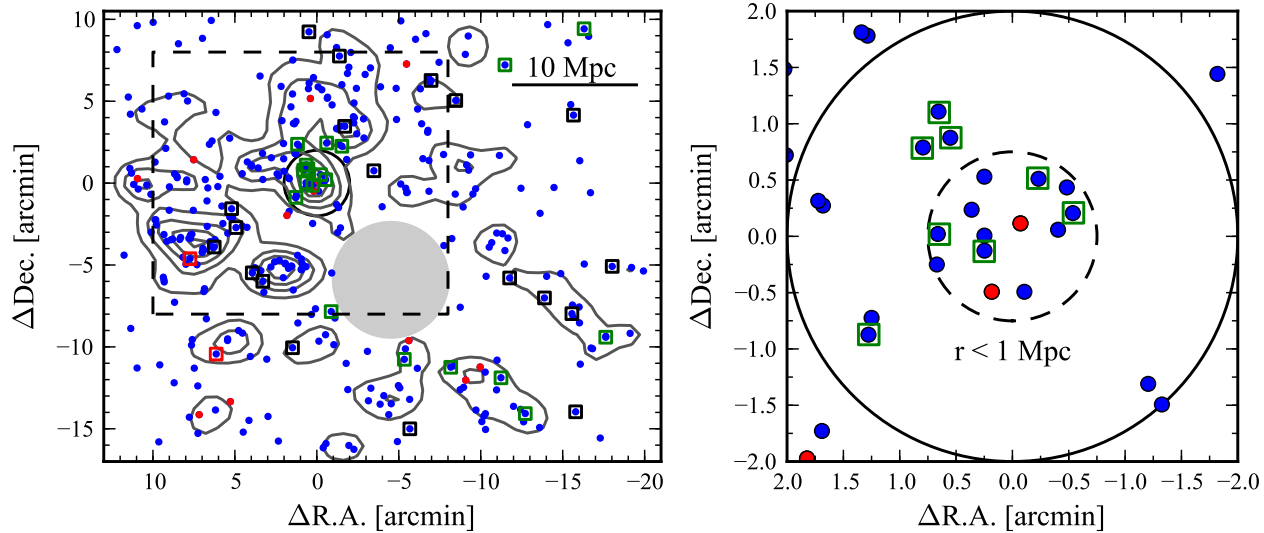}}
\caption{(Left) The 2-D distributions of 352 \oii emitters. Blue and red filled circles show star-forming and red \oii emitters, respectively. Squares indicate the spectroscopically confirmed objects with $1.590\leq z_\mathrm{spec}<1.620$ (black), 1.620$\leq z_\mathrm{spec}\leq 1.630$ (green) and $1.630<z_\mathrm{spec}\leq 1.644$ (red). Contours denote the local number density of \oii emitters. Outside of dash lines is the field region defined in section~4.1. A gray filled circle is a masked region near a bright star.
(Right) Close-up view of the cluster core (r$<2$\arcmin). Dashed line circle shows the radius of r$=$1 Mpc. Note that our NB973 filter can prove about $\pm47$ Mpc (comving) in the redshift direction.}
\label{fig;distribution}
\end{figure*}

It is known that cluster galaxies exhibit a conspicuous ``red sequence'' on the colour-magnitude diagram which are mainly composed of passively evolving galaxies. The red sequence is well recognized even in distant clusters out to $z\sim1$ \citep{1998A&A...334...99K} and it persists at the bright end even in proto-clusters at $z\sim2$ \citep{2007MNRAS.377.1717K}. The red sequence consists of not only passive galaxies but also dusty star-forming galaxies, and we use a $zJK$ diagram to separate them in the same manner as in \cite{2009ApJ...691.1879W}.  Figure \ref{fig;quiescent} shows the $zJK$ diagram for the photo-$z$ selected galaxies. Using the model colour tracks of \cite{1998A&A...334...99K} and the dust reddening vector \citep{2000ApJ...533..682C}, we defined the quiescent galaxies at $z=1.6$ using the three criteria: $z - J > 1.3, J - K < 1.6$ and $z - J > 0.64 (J - K)+0.8$.
In total, 259 quiescent galaxies have been selected from photo-$z$ selected sample.
On the other hand, there are a number of blue galaxies (915) in photo-$z$ selected sample.
It is likely that either their \oii emissions are too weak to be detected as the narrow-band excess or they are outside of the redshift range of the NB973 filter ($z<1.590$ or $z>1.644$). Since the contamination of the latter objects are probably large, we do not use those blue galaxies in our analyses.
Three quiescent galaxies have been spectroscopically confirmed and a spectral fit of the brightest member yields an age of 1.8 Gyr with a mass of $3.8\times10^{11}$M$_\odot$ in Salpeter IMF \citep{2010ApJ...716L.152T}. Because most of galaxies with spectroscopic redshift are line emitters, a follow-up spectroscopy are needed to confirm whether other quiescent galaxies are truly cluster members.

We thus constructed a combined sample of the photo-$z$ selected and mass-limited galaxies at $1.56\leq z_\mathrm{phot}\leq 1.68$, composed of 132 \oii emitters and 259 quiescent galaxies. Table 1 gives a summary of our galaxy samples used in this paper.

\section{Result}

\subsection{Spatial distribution}
\label{sec;spatial}

With the panoramic imaging of the $z=1.62$ cluster with Suprime-Cam, we have revealed a gigantic structure surrounding the cluster traced by \oii emitters for the first time, as shown in Figure \ref{fig;distribution}. The cluster appears to be embedded in a huge filament extending from North to East/South. In particular, there are two dense regions to the East/South and to the South of the cluster, respectively.
Also a filament of relatively dense region extends to the North. They all appear to be associated to the cluster core and constitute a huge structure of about 20 Mpc in comoving scale. 

Many star-forming \oii emitters are located within the projected radius of 1 Mpc in comoving scale from the cluster centre.
This is in stark contrast to the nearby Universe where galaxies in dense environment tend to be in-active red galaxies and star forming galaxies are preferentially located in lower density environments. 
We evaluated the overdenisity of the \oii emitters in the cluster core (r$<$1 Mpc). The average number density of the \oii emitters is 0.24 Mpc$^{-2}$ (352/1472) for the entire region of the survey ($\sim$1472 Mpc$^2$). 
This value is roughly comparable to the number density (0.35 Mpc$^{-2}$) expected from the \oii luminosity function at $z=1.47$ in the general field \citep{2007ApJ...657..738L} because the criterion of the equivalent width in our selection is more stringent than the one used in the previous studies. 
On the other hand, since 13 \oii emitters exist in the cluster core ($\sim$3.142 Mpc$^2$), the number density is 4.1 Mpc$^{-2}$ which is about 17 times larger than that in the entire region. 
The quiescent galaxies show the overdensity by factor of 9.
This clearly indicates that the star formation activity has not been ceased yet even in the core of the $z=1.62$ cluster, and rather the integrated star formation activity in the unit volume is actually higher in the cluster core than in the lower density regions.

Note that our narrow-band filter is probing $\sim$94 Mpc (comving) in the redshift direction, and the 2-D overdensity may not be free from the projection effect.
In order to robustly measure the overdensity, the spectroscopic redshifts would be required.
In fact, it is found from the FMOS spectroscopy that the \oii emitters in the East and the South clumps are located at somewhat different redshifts from the cluster centre (Figure \ref{fig;distribution}).
This difference corresponds to 1000--2000 km s$^{-1}$ in the line of sight velocity or to 15--30 Mpc in the comoving distance.
It is not clear at this stage whether these clumps are gravitationally bound, physically associated systems to the cluster. More intensive spectroscopic follow-up observations are needed to address this issue.

\subsection{Density dependence}
\label{sec;density}

It is widely recognized that the star formation activity of a galaxy is strongly related to the local density of galaxies surrounding it. In the local universe, the star forming activity decreases with increasing local galaxy density \citep{2003ApJ...584..210G,2009ApJ...705L..67P}. This relation holds out to at least $z\sim0.8$ (e.g., \citealt{2001ApJ...562L...9K,2005MNRAS.362..268T}).
At $z\sim1$, it is suggested that SFR-density relation may, in part, be inverted relative to the local relation \citep{2007A&A...468...33E,2008MNRAS.383.1058C}. However, these studies do not include the very dense environment such as rich cluster cores.
In the ClG J2018-0510 cluster at $z=1.62$, \cite{2010ApJ...719L.126T} reported that the relative fraction of star-forming galaxies increases with increasing local density based on the SFRs derived from the 24 $\mu$m fluxes and the SED-fitting. 
Note that they used sample based on only photometric redshift.
On the other hand, \cite{2012ApJ...744...88Q} claimed that the star formation-density relation holds out to at least $z\sim1.8$ although they similarly selected sample on the basis of the photometric redshift. In the case of photo-z selected sample, any results must be interpreted with caution due to a large contamination. Therefore, the reversal of SFR-density relation at $z>1$ is very controversial at this stage.

\begin{table}
\begin{center}
\caption{The numbers of our photo-$z$ selected samples at $1.56\leq z_\mathrm{phot}\leq 1.68$.}
\begin{tabular}{lccc}
\hline
Sample  & Cluster & Field & Total\\
\hline
star-forming \oii emitters &74&50& 124   \\
red \oii emitters &4&4&8 \\
quiescent galaxies &86&173& 259  \\
\hline
combined sample &164&227 & 391 \\
\hline
\end{tabular}
\end{center}
\end{table}

\begin{figure}
\centerline{\includegraphics[width=\linewidth]{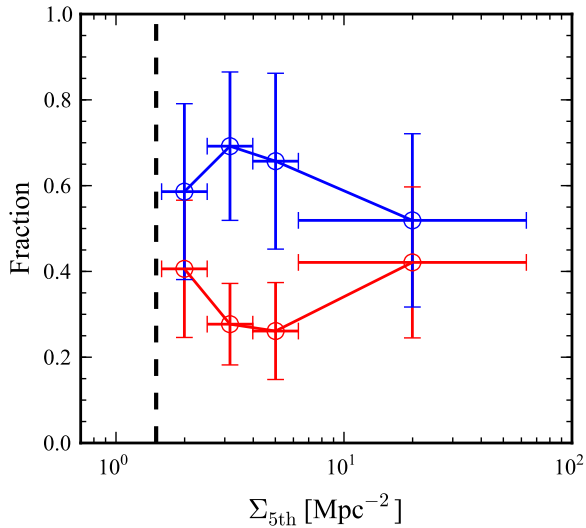}}
\caption{Relative fractions of \oii emitters (blue circles) and quiescent galaxies (red circles) plotted as a function of local density. The local density is calculated from the combined samples of \oii emitters and quiescent galaxies. Dashed line shows the number density in the field region ($\sim 1.5$~Mpc$^{-2}$). In our analysis, we did not use the area below this density. Vertical and horizontal error bars indicate the Poissonian errors and the sizes of each bin, respectively.}
\label{fig;fraction}
\end{figure}

Our panoramic narrow-band survey of the \oii emitters enables us to investigate the environmental dependence over a much larger area hence covering a much wider range in environment. Moreover, due to the great advantage of narrow-band imaging survey combined with photometric redshifts, our \oii emitters are more robust members of star-forming galaxies associated to the cluster even without spectroscopy because we just need to separate out \oii emitters at $z\sim1.6$ among a few other possibilities of different redshifts (corresponding to other lines) based on photometric redshifts.
In order to discuss the environmental effect, we define the local density within the circled area of a radius to the 5th nearest neighbour object ($\Sigma_{\mathrm{5th}}$), by using the combined sample of $z\sim1.6$ galaxies (section 3.3).
And we calculated the fraction of each sample among the combined sample, i.e., $f=N_{\mathrm{each~sample}}/N_{\mathrm{combined~sample}}$ at each bin of the local density.
However, due to the photometric redshift selection, contaminations would dominate the number density in the low density regions.
Therefore, we do not use sample in the field region and concentrated on the region within the dashed rectangle in Figure \ref{fig;distribution}, and trust only the range of local number density that is more than the field density ($\Sigma_{\mathrm{5th,combined}}>1.5$~Mpc$^{-2}$).
The quiescent galaxies sample inevitably contains some foreground or background objects because the photometric redshift range that we adopted ($\Delta z_{\mathrm{phot}}=0.12$) is larger than that of the \oii emitters ($\Delta z_{\mathrm{NB}}=0.054$). We tried to correct for this effect by multiplying $\Delta z_{\mathrm{NB}}/\Delta z_{\mathrm{phot}}$ to the number of quiescent galaxies, assuming conservatively that their redshift distribution is uniform.
As shown in Figure \ref{fig;fraction}, the fraction of star-forming galaxies in the combined sample do not shows a significant dependence on the local density at $z\sim1.6$. It is consistent with a constant value ($\sim$60\%) across different environments within errors. Star-forming activity in the cluster core is very high in this cluster, and the well established star formation--density relation in the local Universe no longer exists.
Although a lot of massive, quiescent galaxies do exist in the high density environment such as the cluster core, the star formation activity has not been ceased yet. Rather the integrated star formation rate per unit volume is actually much higher in the cluster core due to its high over-density of the emitters.

\begin{figure}
\centerline{\includegraphics[width=\linewidth]{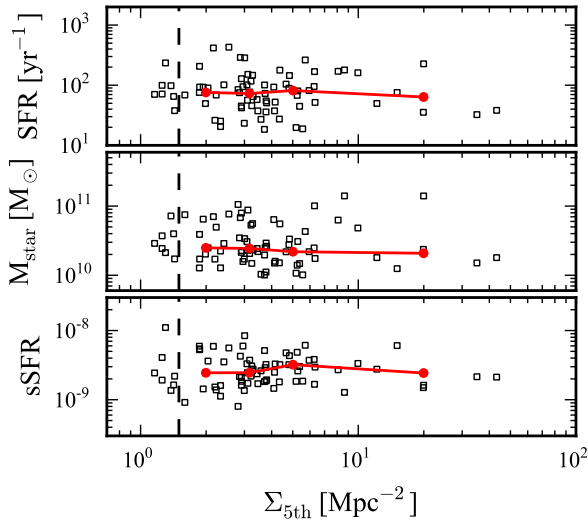}}
\caption{SFRs (top), stellar masses (middle) and specific SFRs (bottom) for individual \oii emitters, plotted against the local density. The local density is calculated from the combined samples of \oii emitters and quiescent galaxies. Red circles show the median value in each density bin.}
\label{fig;property}
\end{figure}

In order to quantitatively compare the properties of the individual \oii emitters as a function of environment, we estimate stellar mass, SFR and specific SFR (SSFR=SFR/M$_\mathrm{star}$) for the \oii emitters. 
\oii emission line is widely used as a good SFR indicator but strongly depends on metallicity and dust-extinction. \oii-SFR relation are recently being calibrated by using lower redshift samples \citep{2010MNRAS.405.2594G,2012MNRAS.420.1926S}. We used the \oii-SFR calibration given by \cite{2012MNRAS.420.1926S}, which is described as a function of stellar mass.
The results are shown in Figure \ref{fig;property} as a function of local density. We do not find any significant environmental dependence in all of the individual quantities of the \oii emitters. In fact, the median values of all the three quantities are almost constant with local density.
\cite{2011PASJ...63S.437T} claimed that the averaged star formation activity is rapidly declined from $z$=2.5 to $z$=0.8 in high density regions such as clusters while it is only gradually declined in the low density regions. At some point at high redshifts, we may expect that the averaged star formation rate of galaxies in cluster cores should exceed that in the general field probably because galaxy formation processes (such as gas cooling and mergers) are accelerated in dense environments. The fact that we observe just a comparable level of star formation activity in the \oii emitters irrespective of environment may suggest that, at $z=1.62$, the \oii emitters in the high density regions are just in the transition phase from a star-bursting mode to a quiescent mode as a result of environmental effect.

\section{Summary}

We have conducted a narrow-band survey of \oii emitters in and around the ClG J0218.3--0510 cluster at $z=1.6$, using Suprime-Cam on Subaru. 
The observation with $z_\mathrm{R}$ filter was newly carried out to measure the continuum at the same wavelength as the narrow-band filter (NB973).
We combined these data with multi broad-band data ($B,V,R,i',z',J,H,K,3.6~\mu$m,$4.5~\mu$m) to create the NB973-detected and K-detected photometric catalogues.
The photometric redshifts were determined with 11-bands including $z_\mathrm{R}$-band. By comparing the spectroscopic members and our \oii emitter sample, we estimated that the accuracy of our photometric redshifts is $\sigma _z=0.05$ at $z=1.6$. Our main results with these samples are summarized as follows.

\begin{enumerate}

\item On the basis of narrow-band excesses and photometric redshifts, our survey provides a sample of 352 \oii emitters over a 830 arcmin$^2$ area. 
Our very recent FMOS near-infrared spectroscopic observations have confirmed 31 \oii emitters at $z\sim1.6$ by the presence of \ha or \oiii lines at the expected wavelengths.
The \oii emitters constitute a large scale structure at $z=1.62$ in which the ClG J0218.3--0510 cluster is embedded. Also, we find that many star-forming \oii emitters are located even in the cluster core (r$<$1~Mpc in the comoving scale) and in the surrounding clumps, and show a high overdensity by a factor of 17 compared to the entire region. This suggests that the integrated star formation activity per unit volume is activated in such regions.

\vspace{1mm}

\item The galaxies with $1.56 \leq z_{\mathrm{phot}}\leq 1.68$ show a clear bimodal distribution in the $zJK$ diagram, namely, quiescent and star-forming galaxies. We selected total of 259 quiescent galaxies on the basis of the population synthesis model and the extinction law by dust. 

\vspace{1mm}

\item We calculated the local number density of galaxies in the photo-$z$ selected and mass-limited sample to examine the environmental dependence of the star-formation activity. 
We obtain a large fraction of \oii emitters even in the cluster core, showing that the star forming activity in the cluster core is elevated substantially compared to the local clusters where there are little star-forming galaxies in their cores.
There is no longer a environmental dependence in the relative fraction of \oii emitters in the combined sample, and the well known SFR-density relation in the present-day Universe no longer exists within errors.
Furthermore, the properties of the individual \oii emitters, such as star formation rates, stellar masses and specific star formation rates, do not depend on the local density, either.
These results suggest that the \oii emitters in the high density regions are just in the transition phase from a star-bursting mode to a quiescent mode due to some environmental effects and the star formation rates in these systems may be rapidly declining.

\end{enumerate}

Note that the results presented in this paper for the environmental effect of galaxies in and around a $z$=1.62 cluster may not necessarily apply to all the clusters at the same epoch of the Universe, and the results may well depend on the degree of matureness of clusters even at the same redshift.
For this reason, we require a systematic studies of distant clusters and general fields to construct a more general picture, and the MAHALO-Subaru project (Kodama et al. in prep) will provide us more comprehensive views of galaxy evolution at its most active phase in the Universe.

\section*{Acknowledgments}

This paper is based on data collected at Subaru Telescope, which is operated by the National Astronomical Observatory of Japan. We thank the Subaru telescope staff for their help in the observation. 
We thank the anonymous referee who gave us many useful comments, which improved the paper.
T.K. acknowledges the financial support in part by a Grant-in-Aid for the Scientific Research (No.\, 21340045) by the Japanese Ministry of Education, Culture, Sports, Science and Technology.
Y.K. acknowledge the support from the Japan Society for the Promotion of Science (JSPS) through JSPS research fellowships for young scientists.  

\bibliographystyle{mn2e.bst}
\bibliography{tadaki_2011b}

\begin{thebibliography}{}

\bibitem[\protect\citeauthoryear{{Baldwin}, {Phillips} \&
  {Terlevich}}{{Baldwin} et~al.}{1981}]{1981PASP...93....5B}
{Baldwin} J.~A.,  {Phillips} M.~M.,    {Terlevich} R.,  1981, \pasp, 93, 5

\bibitem[\protect\citeauthoryear{{Bauer}, {Gr{\"u}tzbauch}, {J{\o}rgensen},
  {Varela} \& {Bergmann}}{{Bauer} et~al.}{2011}]{2011MNRAS.411.2009B}
{Bauer} A.~E.,  {Gr{\"u}tzbauch} R.,  {J{\o}rgensen} I.,  {Varela} J.,
  {Bergmann} M.,  2011, \mnras, 411, 2009

\bibitem[\protect\citeauthoryear{{Bertin} \& {Arnouts}}{{Bertin} \&
  {Arnouts}}{1996}]{1996A&AS..117..393B}
{Bertin} E.,  {Arnouts} S.,  1996, \aaps, 117, 393

\bibitem[\protect\citeauthoryear{{Brammer}, {van Dokkum} \& {Coppi}}{{Brammer}
  et~al.}{2008}]{2008ApJ...686.1503B}
{Brammer} G.~B.,  {van Dokkum} P.~G.,    {Coppi} P.,  2008, \apj, 686, 1503

\bibitem[\protect\citeauthoryear{{Brodwin}, {Stern}, {Vikhlinin}, {Stanford},
  {Gonzalez}, {Eisenhardt}, {Ashby}, {Bautz}, {Dey}, {Forman}, {Gettings},
  {Hickox}, {Jannuzi}, {Jones}, {Mancone}, {Miller}, {Moustakas}, {Ruel},
  {Snyder} \& {Zeimann}}{{Brodwin} et~al.}{2011}]{2011ApJ...732...33B}
{Brodwin} M. et~al., 2011, \apj, 732, 33

\bibitem[\protect\citeauthoryear{{Bunker}, {Warren}, {Hewett} \&
  {Clements}}{{Bunker} et~al.}{1995}]{1995MNRAS.273..513B}
{Bunker} A.~J.,  {Warren} S.~J.,  {Hewett} P.~C.,    {Clements} D.~L.,  1995,
  \mnras, 273, 513

\bibitem[\protect\citeauthoryear{{Calzetti}, {Armus}, {Bohlin}, {Kinney},
  {Koornneef} \& {Storchi-Bergmann}}{{Calzetti}
  et~al.}{2000}]{2000ApJ...533..682C}
{Calzetti} D.,  {Armus} L.,  {Bohlin} R.~C.,  {Kinney} A.~L.,  {Koornneef} J.,
    {Storchi-Bergmann} T.,  2000, \apj, 533, 682

\bibitem[\protect\citeauthoryear{{Cooper}, {Newman}, {Weiner}, {Yan},
  {Willmer}, {Bundy}, {Coil}, {Conselice}, {Davis}, {Faber}, {Gerke},
  {Guhathakurta}, {Koo} \& {Noeske}}{{Cooper}
  et~al.}{2008}]{2008MNRAS.383.1058C}
{Cooper} M.~C. et~al., 2008, \mnras, 383, 1058

\bibitem[\protect\citeauthoryear{{Dressler}}{{Dressler}}{1980}]{1980ApJ...236.%
.351D}
{Dressler} A.,  1980, \apj, 236, 351

\bibitem[\protect\citeauthoryear{{Dressler}, {Oemler} Jr., {Couch}, {Smail},
  {Ellis}, {Barger}, {Butcher}, {Poggianti} \& {Sharples}}{{Dressler}
  et~al.}{1997}]{1997ApJ...490..577D}
{Dressler} A. et~al., 1997,
  \apj, 490, 577

\bibitem[\protect\citeauthoryear{{Elbaz}, {Daddi}, {Le Borgne}, {Dickinson},
  {Alexander}, {Chary}, {Starck}, {Brandt}, {Kitzbichler}, {MacDonald},
  {Nonino}, {Popesso}, {Stern} \& {Vanzella}}{{Elbaz}
  et~al.}{2007}]{2007A&A...468...33E}
{Elbaz} D. et~al., 2007, \aap,
  468, 33

\bibitem[\protect\citeauthoryear{{Fassbender}, {Nastasi}, {B{\"o}hringer}, {{\v
  S}uhada}, {Santos}, {Rosati}, {Pierini}, {M{\"u}hlegger}, {Quintana},
  {Schwope}, {Lamer}, {de Hoon}, {Kohnert}, {Pratt} \& {Mohr}}{{Fassbender}
  et~al.}{2011}]{2011A&A...527L..10F}
{Fassbender} R. et~al.,  2011, \aap, 527, L10+

\bibitem[\protect\citeauthoryear{{Furusawa}, {Kosugi}, {Akiyama}, {Takata},
  {Sekiguchi}, {Tanaka}, {Iwata}, {Kajisawa}, {Yasuda}, {Doi}, {Ouchi},
  {Simpson}, {Shimasaku}, {Yamada}, {Furusawa}, {Morokuma}, {Ishida}, {Aoki},
  {Fuse}, {Imanishi}, {Iye}}{Furusawa et~al.}{2008}]{2008ApJS..176....1F}
{Furusawa} H. et~al., 2008,
  \apjs, 176, 1
  
\bibitem[\protect\citeauthoryear{{Gilbank}, {Baldry}, {Balogh}, {Glazebrook} \&
  {Bower}}{{Gilbank} et~al.}{2010}]{2010MNRAS.405.2594G}
{Gilbank} D.~G.,  {Baldry} I.~K.,  {Balogh} M.~L.,  {Glazebrook} K.,    {Bower}
  R.~G.,  2010, \mnras, 405, 2594

\bibitem[\protect\citeauthoryear{{Gobat}, {Daddi}, {Onodera}, {Finoguenov},
  {Renzini}, {Arimoto}, {Bouwens}, {Brusa}, {Chary}, {Cimatti}, {Dickinson},
  {Kong} \& {Mignoli}}{{Gobat} et~al.}{2011}]{2011A&A...526A.133G}
{Gobat} R. et~al., 2011, \aap, 526, A133+

\bibitem[\protect\citeauthoryear{{G{\'o}mez}, {Nichol}, {Miller}, {Balogh},
  {Goto}, {Zabludoff}, {Romer}, {Bernardi}, {Sheth}, {Hopkins}, {Castander},
  {Connolly}, {Schneider}, {Brinkmann}, {Lamb}, {SubbaRao} \&
  {York}}{{G{\'o}mez} et~al.}{2003}]{2003ApJ...584..210G}
{G{\'o}mez} P.~L. et~al., 2003, \apj,
  584, 210

\bibitem[\protect\citeauthoryear{{Hayashi}, {Kodama}, {Koyama}, {Tadaki} \&
  {Tanaka}}{{Hayashi} et~al.}{2011}]{2011MNRAS.415.2670H}
{Hayashi} M.,  {Kodama} T.,  {Koyama} Y.,  {Tadaki} K.-I.,    {Tanaka} I.,
  2011, \mnras, 415, 2670

\bibitem[\protect\citeauthoryear{{Hayashi}, {Kodama}, {Koyama}, {Tanaka},
  {Shimasaku} \& {Okamura}}{{Hayashi} et~al.}{2010}]{2010MNRAS.402.1980H}
{Hayashi} M.,  {Kodama} T.,  {Koyama} Y.,  {Tanaka} I.,  {Shimasaku} K.,
  {Okamura} S.,  2010, \mnras, 402, 1980

\bibitem[\protect\citeauthoryear{{Henry}, {Salvato}, {Finoguenov}, {Bouche},
  {Brunner}, {Burwitz}, {Buschkamp}, {Egami}, {F{\"o}rster-Schreiber},
  {Fotopoulou}, {Genzel}, {Hasinger}, {Mainieri}, {Rovilos} \&
  {Szokoly}}{{Henry} et~al.}{2010}]{2010ApJ...725..615H}
{Henry} J.~P. et~al.,  2010, \apj, 725, 615

\bibitem[\protect\citeauthoryear{{Hewett}, {Warren}, {Leggett} \&
  {Hodgkin}}{{Hewett} et~al.}{2006}]{2006MNRAS.367..454H}
{Hewett} P.~C.,  {Warren} S.~J.,  {Leggett} S.~K.,    {Hodgkin} S.~T.,  2006,
  \mnras, 367, 454

\bibitem[\protect\citeauthoryear{{Hodgkin}, {Irwin}, {Hewett} \&
  {Warren}}{{Hodgkin} et~al.}{2009}]{2009MNRAS.394..675H}
{Hodgkin} S.~T.,  {Irwin} M.~J.,  {Hewett} P.~C.,    {Warren} S.~J.,  2009,
  \mnras, 394, 675

\bibitem[\protect\citeauthoryear{{Iwamuro}, {Moritani}, {Yabe}, {Sumiyoshi},
  {Kawate}, {Tamura}, {Akiyama}, {Kimura}, {Takato}, {Tait}, {Ohta}, {Totani},
  {Suzuki} \& {Tonegawa}}{{Iwamuro} et~al.}{2011}]{2011arXiv1111.6746I}
{Iwamuro} F. et~al., 2011, ArXiv e-prints

\bibitem[\protect\citeauthoryear{{Kauffmann}, {Heckman}, {Tremonti},
  {Brinchmann}, {Charlot}, {White}, {Ridgway}, {Brinkmann}, {Fukugita}, {Hall},
  {Ivezi{\'c}}, {Richards} \& {Schneider}}{{Kauffmann}
  et~al.}{2003}]{2003MNRAS.346.1055K}
{Kauffmann} G. et~al., 2003, \mnras, 346, 1055

\bibitem[\protect\citeauthoryear{{Kauffmann}, {White}, {Heckman}, {M{\'e}nard},
  {Brinchmann}, {Charlot}, {Tremonti} \& {Brinkmann}}{{Kauffmann}
  et~al.}{2004}]{2004MNRAS.353..713K}
{Kauffmann} G.,  {White} S.~D.~M.,  {Heckman} T.~M.,  {M{\'e}nard} B.,
  {Brinchmann} J.,  {Charlot} S.,  {Tremonti} C.,    {Brinkmann} J.,  2004,
  \mnras, 353, 713

\bibitem[\protect\citeauthoryear{{Kennicutt}
  Jr.}{{Kennicutt}}{1998}]{1998ARA&A..36..189K}
{Kennicutt} Jr. R.~C.,  1998, \araa, 36, 189

\bibitem[\protect\citeauthoryear{{Kimura}, {Maihara}, {Iwamuro}, {Akiyama},
  {Tamura}, {Dalton}, {Takato}, {Tait}, {Ohta}, {Eto}, {Mochida}, {Elms},
  {Kawate}, {Kurakami}, {Moritani}, {Noumaru}, {Ohshima}, {Sumiyoshi}, {Yabe},
  {Brzeski}}{{Kimura} M. et~al.}{2010}]{2010PASJ...62.1135K}
{Kimura} M. et~al.,  2010, \pasj, 62, 1135

\bibitem[\protect\citeauthoryear{{Kodama}, {Arimoto}, {Barger} \&
  {Arag'on-Salamanca}}{{Kodama} et~al.}{1998}]{1998A&A...334...99K}
{Kodama} T.,  {Arimoto} N.,  {Barger} A.~J.,    {Arag'on-Salamanca} A.,  1998,
  \aap, 334, 99

\bibitem[\protect\citeauthoryear{{Kodama}, {Bell} \& {Bower}}{{Kodama}
  et~al.}{1999}]{1999MNRAS.302..152K}
{Kodama} T.,  {Bell} E.~F.,    {Bower} R.~G.,  1999, \mnras, 302, 152

\bibitem[\protect\citeauthoryear{{Kodama}, {Smail}, {Nakata}, {Okamura} \&
  {Bower}}{{Kodama} et~al.}{2001}]{2001ApJ...562L...9K}
{Kodama} T.,  {Smail} I.,  {Nakata} F.,  {Okamura} S.,    {Bower} R.~G.,  2001,
  \apjl, 562, L9

\bibitem[\protect\citeauthoryear{{Kodama}, {Tanaka}, {Kajisawa}, {Kurk},
  {Venemans}, {De Breuck}, {Vernet} \& {Lidman}}{{Kodama}
  et~al.}{2007}]{2007MNRAS.377.1717K}
{Kodama} T.,  {Tanaka} I.,  {Kajisawa} M.,  {Kurk} J.,  {Venemans} B.,  {De
  Breuck} C.,  {Vernet} J.,    {Lidman} C.,  2007, \mnras, 377, 1717

\bibitem[\protect\citeauthoryear{{Koyama}, {Kodama}, {Nakata}, {Shimasaku} \&
  {Okamura}}{{Koyama} et~al.}{2011}]{2011ApJ...734...66K}
{Koyama} Y.,  {Kodama} T.,  {Nakata} F.,  {Shimasaku} K.,    {Okamura} S.,
  2011, \apj, 734, 66

\bibitem[\protect\citeauthoryear{{Koyama}, {Kodama}, {Shimasaku}, {Hayashi},
  {Okamura}, {Tanaka} \& {Tokoku}}{{Koyama} et~al.}{2010}]{2010MNRAS.403.1611K}
{Koyama} Y.,  {Kodama} T.,  {Shimasaku} K.,  {Hayashi} M.,  {Okamura} S.,
  {Tanaka} I.,    {Tokoku} C.,  2010, \mnras, 403, 1611

\bibitem[\protect\citeauthoryear{{Kron}}{{Kron}}{1980}]{1980ApJS...43..305K}
{Kron} R.~G.,  1980, \apjs, 43, 305

\bibitem[\protect\citeauthoryear{{Lawrence}, {Warren}, {Almaini}, {Edge},
  {Hambly}, {Jameson}, {Lucas}, {Casali}, {Adamson}, {Dye}, {Emerson},
  {Foucaud}, {Hewett}, {Hirst}, {Hodgkin}, {Irwin}, {Lodieu}, {McMahon},
  {Simpson}, {Smail}, {Mortlock} \& {Folger}}{{Lawrence} et~al.}{2007}]{2007MNRAS.379.1599L}
{Lawrence} A. et~al., 2007, \mnras, 379, 1599

\bibitem[\protect\citeauthoryear{{Lemaux}, {Lubin}, {Shapley}, {Kocevski},
  {Gal} \& {Squires}}{{Lemaux} et~al.}{2010}]{2010ApJ...716..970L}
{Lemaux} B.~C.,  {Lubin} L.~M.,  {Shapley} A.,  {Kocevski} D.,  {Gal} R.~R.,
  {Squires} G.~K., 2010, \apj, 716, 970

\bibitem[\protect\citeauthoryear{{Lewis}, {Balogh}, {De Propris}, {Couch},
  {Bower}, {Offer}, {Bland-Hawthorn}, {Baldry}, {Baugh}, {Bridges}, {Cannon},
  {Cole}, {Colless}, {Collins}, {Cross}, {Dalton}, {Driver}, {Efstathiou},
  {Ellis}, {Frenk}, {Glazebrook}}{{Lewis} I. et~al.}{2002}]{2002MNRAS.334..673L}
{Lewis} I. et~al., 2002, \mnras, 334, 673

\bibitem[\protect\citeauthoryear{{Ly}, {Malkan}, {Kashikawa}, {Shimasaku},
  {Doi}, {Nagao}, {Iye}, {Kodama}, {Morokuma} \& {Motohara}}{{Ly}
  et~al.}{2007}]{2007ApJ...657..738L}
{Ly} C. et~al.,  2007, \apj,
  657, 738

\bibitem[\protect\citeauthoryear{{Miyazaki}, {Komiyama}, {Sekiguchi},
  {Okamura}, {Doi}, {Furusawa}, {Hamabe}, {Imi}, {Kimura}, {Nakata}, {Okada},
  {Ouchi}, {Shimasaku}, {Yagi} \& {Yasuda}}{{Miyazaki}
  et~al.}{2002}]{2002PASJ...54..833M}
{Miyazaki} S. et~al., 2002, \pasj,
  54, 833

\bibitem[\protect\citeauthoryear{{Muzzin}, {Wilson}, {Yee}, {Gilbank},
  {Hoekstra}, {Demarco}, {Balogh}, {van Dokkum}, {Franx}, {Ellingson}, {Hicks},
  {Nantais}, {Noble}, {Lacy}, {Lidman}, {Rettura}, {Surace} \& {Webb}}{{Muzzin}
  et~al.}{2011}]{2011arXiv1112.3655M}
{Muzzin} A. et~al., 2011, ArXiv e-prints

\bibitem[\protect\citeauthoryear{{Ota}, {Iye}, {Kashikawa}, {Shimasaku},
  {Ouchi}, {Totani}, {Kobayashi}, {Nagashima}, {Harayama}, {Kodaka},
  {Morokuma}, {Furusawa}, {Tajitsu} \& {Hattori}}{{Ota}
  et~al.}{2010}]{2010ApJ...722..803O}
{Ota} K. et~al., 2010, \apj,
  722, 803

\bibitem[\protect\citeauthoryear{{Ouchi}, {Shimasaku}, {Okamura}, {Furusawa},
  {Kashikawa}, {Ota}, {Doi}, {Hamabe}, {Kimura}, {Komiyama}, {Miyazaki},
  {Miyazaki}, {Nakata}, {Sekiguchi}, {Yagi} \& {Yasuda}}{{Ouchi}
  et~al.}{2004}]{2004ApJ...611..660O}
{Ouchi} M. et~al., 2004, \apj, 611, 660

\bibitem[\protect\citeauthoryear{{Papovich}, {Momcheva}, {Willmer},
  {Finkelstein}, {Finkelstein}, {Tran}, {Brodwin}, {Dunlop}, {Farrah}, {Khan},
  {Lotz}, {McCarthy}, {McLure}, {Rieke}, {Rudnick}, {Sivanandam}, {Pacaud} \&
  {Pierre}}{{Papovich} et~al.}{2010}]{2010ApJ...716.1503P}
{Papovich} C. et~al., 2010, \apj,
  716, 1503

\bibitem[\protect\citeauthoryear{{Patel}, {Holden}, {Kelson}, {Illingworth} \&
  {Franx}}{{Patel} et~al.}{2009}]{2009ApJ...705L..67P}
{Patel} S.~G.,  {Holden} B.~P.,  {Kelson} D.~D.,  {Illingworth} G.~D.,
  {Franx} M., 2009, \apjl, 705, L67

\bibitem[\protect\citeauthoryear{{Postman} \& {Geller}}{{Postman} \&
  {Geller}}{1984}]{1984ApJ...281...95P}
{Postman} M.,  {Geller} M.~J.,  1984, \apj, 281, 95

\bibitem[\protect\citeauthoryear{{Quadri}, {Williams}, {Franx} \&
  {Hildebrandt}}{{Quadri} et~al.}{2012}]{2012ApJ...744...88Q}
{Quadri} R.~F.,  {Williams} R.~J.,  {Franx} M.,    {Hildebrandt} H.,  2012,
  \apj, 744, 88

\bibitem[\protect\citeauthoryear{{Simpson}, {Rawlings}, {Ivison}, {Akiyama},
  {Almaini}, {Bradshaw}, {Chapman}, {Chuter}, {Croom}, {Dunlop}, {Foucaud} \&
  {Hartley}}{{Simpson} et~al.}{2012}]{2012MNRAS.tmp.2493S}
{Simpson} C. et~al., 2012, \mnras, p.~2493

\bibitem[\protect\citeauthoryear{{Skrutskie}, {Cutri}, {Stiening}, {Weinberg},
  {Schneider}, {Carpenter}, {Beichman}, {Capps}, {Chester}, {Elias}, {Huchra},
  {Liebert}, {Lonsdale}, {Monet}, {Price}, {Seitzer}, {Jarrett}}{{Skrutskie} et~al.}{2006}]{2006AJ...131.1163S}
{Skrutskie} M.~F. et al., 2006, \aj, 131, 1163

\bibitem[\protect\citeauthoryear{{Smail}, {Sharp}, {Swinbank}, {Akiyama},
  {Ueda}, {Foucaud}, {Almaini} \& {Croom}}{{Smail}
  et~al.}{2008}]{2008MNRAS.389..407S}
{Smail} I.,  {Sharp} R.,  {Swinbank} A.~M.,  {Akiyama} M.,  {Ueda} Y.,
  {Foucaud} S.,  {Almaini} O.,    {Croom} S.,  2008, \mnras, 389, 407

\bibitem[\protect\citeauthoryear{{Sobral}, {Best}, {Geach}, {Smail}, {Kurk},
  {Cirasuolo}, {Casali}, {Ivison}, {Coppin} \& {Dalton}}{{Sobral}
  et~al.}{2009}]{2009MNRAS.398...75S}
{Sobral} D. et~al., 2009, \mnras, 398, 75

\bibitem[\protect\citeauthoryear{{Sobral}, {Best}, {Matsuda}, {Smail}, {Geach}
  \& {Cirasuolo}}{{Sobral} et~al.}{2012}]{2012MNRAS.420.1926S}
{Sobral} D.,  {Best} P.~N.,  {Matsuda} Y.,  {Smail} I.,  {Geach} J.~E.,
  {Cirasuolo} M.,  2012, \mnras, 420, 1926

\bibitem[\protect\citeauthoryear{{Sobral}, {Best}, {Smail}, {Geach},
  {Cirasuolo}, {Garn} \& {Dalton}}{{Sobral} et~al.}{2011}]{2011MNRAS.411..675S}
{Sobral} D.,  {Best} P.~N.,  {Smail} I.,  {Geach} J.~E.,  {Cirasuolo} M.,
  {Garn} T.,    {Dalton} G.~B.,  2011, \mnras, 411, 675

\bibitem[\protect\citeauthoryear{{Tadaki}, {Kodama}, {Koyama}, {Hayashi},
  {Tanaka} \& {Tokoku}}{{Tadaki} et~al.}{2011}]{2011PASJ...63S.437T}
{Tadaki} K.-I.,  {Kodama} T.,  {Koyama} Y.,  {Hayashi} M.,  {Tanaka} I.,
  {Tokoku} C.,  2011, \pasj, 63, 437

\bibitem[\protect\citeauthoryear{{Tanaka}, {Finoguenov}, {Lilly}, {Bolzonella},
  {Carollo}, {Contini}, {Iovino}, {Kneib}, {Lamareille}, {Le Fevre},
  {Mainieri}, {Presotto}, {Renzini}, {Scodeggio}, {Silverman}, {Zamorani},
  {Bardelli}, {Bongiorno}}{{Tanaka} M. et~al.}{2011}]{2011arXiv1110.0979T}
{Tanaka} M. et~al., 2011, ArXiv e-prints

\bibitem[\protect\citeauthoryear{{Tanaka}, {Finoguenov} \& {Ueda}}{{Tanaka}
  et~al.}{2010}]{2010ApJ...716L.152T}
{Tanaka} M.,  {Finoguenov} A.,    {Ueda} Y.,  2010, \apjl, 716, L152

\bibitem[\protect\citeauthoryear{{Tanaka}, {Kodama}, {Arimoto}, {Okamura},
  {Umetsu}, {Shimasaku}, {Tanaka} \& {Yamada}}{{Tanaka}
  et~al.}{2005}]{2005MNRAS.362..268T}
{Tanaka} M.,  {Kodama} T.,  {Arimoto} N.,  {Okamura} S.,  {Umetsu} K.,
  {Shimasaku} K.,  {Tanaka} I.,    {Yamada} T.,  2005, \mnras, 362, 268

\bibitem[\protect\citeauthoryear{{Tran}, {Papovich}, {Saintonge}, {Brodwin},
  {Dunlop}, {Farrah}, {Finkelstein}, {Finkelstein}, {Lotz}, {McLure},
  {Momcheva} \& {Willmer}}{{Tran} et~al.}{2010}]{2010ApJ...719L.126T}
{Tran} K. et~al., 2010, \apjl, 719,
  L126

\bibitem[\protect\citeauthoryear{{Ueda}, {Watson}, {Stewart}, {Akiyama},
  {Schwope}, {Lamer}, {Ebrero}, {Carrera}, {Sekiguchi}, {Yamada}, {Simpson},
  {Hasinger} \& {Mateos}}{{Ueda} et~al.}{2008}]{2008ApJS..179..124U}
{Ueda} Y. et~al.,  2008, \apjs, 179, 124

\bibitem[\protect\citeauthoryear{{Whitmore}, {Gilmore} \& {Jones}}{{Whitmore}
  et~al.}{1993}]{1993ApJ...407..489W}
{Whitmore} B.~C.,  {Gilmore} D.~M.,    {Jones} C.,  1993, \apj, 407, 489

\bibitem[\protect\citeauthoryear{{Williams}, {Quadri}, {Franx}, {van Dokkum} \&
  {Labb{\'e}}}{{Williams} et~al.}{2009}]{2009ApJ...691.1879W}
{Williams} R.~J.,  {Quadri} R.~F.,  {Franx} M.,  {van Dokkum} P.,
  {Labb{\'e}} I.,  2009, \apj, 691, 1879

\bibitem[\protect\citeauthoryear{{Yagi}, {Kashikawa}, {Sekiguchi}, {Doi},
  {Yasuda}, {Shimasaku} \& {Okamura}}{{Yagi}
  et~al.}{2002}]{2002AJ....123...66Y}
{Yagi} M.,  {Kashikawa} N.,  {Sekiguchi} M.,  {Doi} M.,  {Yasuda} N.,
  {Shimasaku} K.,    {Okamura} S.,  2002, \aj, 123, 66

\bibitem[\protect\citeauthoryear{{Yan}, {Newman}, {Faber}, {Konidaris}, {Koo}
  \& {Davis}}{{Yan} et~al.}{2006}]{2006ApJ...648..281Y}
{Yan} R.,  {Newman} J.~A.,  {Faber} S.~M.,  {Konidaris} N.,  {Koo} D.,
  {Davis} M.,  2006, \apj, 648, 281

\end{thebibliography}

\label{lastpage}

\end{document}